\documentclass[submission,copyright,creativecommons]{eptcs}

\usepackage{breakurl}             
\usepackage{underscore}           
\usepackage{amssymb}
\usepackage{listings}
\usepackage{indentfirst}
\usepackage{verbatim}
\usepackage{amsmath, amsthm, amssymb}
\usepackage{graphicx}
\usepackage{url}
\usepackage{hyperref}
\usepackage{xspace}
\usepackage{placeins}

\lstdefinelanguage{scheme}{
keywords={define, conde, fresh},
sensitive=true,
commentstyle=\scriptsize\rmfamily,
keywordstyle=\ttfamily\underbar,
identifierstyle=\ttfamily,
basewidth={0.5em,0.5em},
columns=fixed,
fontadjust=true,
literate={==}{{$\equiv$}}1
}

\lstdefinelanguage{ocaml}{
keywords={fresh, conde, let, begin, end, in, match, type, and, fun, function, try, with, when, class,
object, method, of, rec, repeat, until, while, not, do, done, as, val, inherit,
new, module, sig, deriving, datatype, struct, if, then, else, open, private, virtual, include, @type},
sensitive=true,
commentstyle=\small\itshape\ttfamily,
keywordstyle=\ttfamily\underbar,
identifierstyle=\ttfamily,
basewidth={0.5em,0.5em},
columns=fixed,
fontadjust=true,
literate={->}{{$\to\;\;$}}3 {===}{{$\equiv$}}3 {=/=}{{$\not\equiv$}}3 {|>}{{$\triangleright$}}3,
morecomment=[s]{(*}{*)}
}

\newcommand{\miniKanren}{miniKanren\xspace}

\title{Typed Embedding of a Relational Language in OCaml}
\author{Dmitrii Kosarev
\institute{Saint Petersburg State University\\ Saint Petersburg, Russia}
\email{Dmitrii.Kosarev@protonmail.ch}
\and
Dmitry Boulytchev
\institute{Saint Petersburg State University\\ Saint Petersburg, Russia}
\email{dboulytchev@math.spbu.ru}
}
\def\titlerunning{Typed Embedding of a Relational Language in OCaml}
\def\authorrunning{Dmitrii Kosarev \& Dmitry Boulytchev}
\begin{document}
\maketitle

\begin{abstract}
We present an implementation of the relational programming language \miniKanren as a set
of combinators and syntax extensions for OCaml. The key feature of our approach is
\emph{polymorphic unification}, which can be used to unify data structures of arbitrary types.
In addition we provide a useful generic programming pattern to systematically develop relational
specifications in a typed manner, and address the problem of integration of relational subsystems into
functional applications.
\end{abstract}

\section{Introduction}
\label{intro}

Relational programming~\cite{TRS} is an attractive technique, based on the idea
of constructing programs as relations.  As a result, relational programs can be
``queried'' in various ``directions'', making it possible, for example, to simulate
reversed execution. Apart from being interesting from a purely theoretical standpoint,
this approach may have a practical value: some problems look much simpler
when considered as queries to some relational specification~\cite{WillThesis}. There are a
number of appealing examples confirming this observation: a type checker
for simply typed lambda calculus (and, at the same time, a type inferencer and solver
for the inhabitation problem), an interpreter (capable of generating ``quines''~---
programs producing themselves as a result)~\cite{Untagged}, list sorting (capable of
producing all permutations), etc.

Many logic programming languages, such as Prolog, Mercury~\cite{MercuryFirstPaper},
or Curry~\cite{CurryFirstPaper} to some extent
can be considered relational. We have chosen \miniKanren\footnote{\url{http://minikanren.org}}
as a model language, because it was specifically designed as a relational DSL, embedded in Scheme/Racket.
Being rather a minimalistic language, which can be implemented with just a few data structures and
combinators~\cite{MicroKanren, MuKanrenNew}, \miniKanren found its way into dozens of host languages, including Scala, Haskell and Standard ML.
The paradigm behind \miniKanren can be described as ``lightweight logic programming''\footnote{An in-depth comparison of \miniKanren
and Prolog can be found here: \url{http://minikanren.org/minikanren-and-prolog.html}}.

This paper addresses the problem of embedding \miniKanren into OCaml\footnote{\url{http://ocaml.org}}~--- a statically-typed functional language with
a rich type system. A statically-typed implementation would bring us a number of benefits. First, as always,
we expect typing to provide a certain level of correctness guarantees, ruling out some pathological programs, which
otherwise would provide pathological results. In the context of relational programming, however, typing would additionally
help us to interpret the results of queries. Often an answer to a relational query contains a number of
free variables, which are supposed to take arbitrary values. In the typed case these variables become typed,
facilitating the understanding of the answers, especially those with multiple free variables. Next, a number of \miniKanren
applications require additional constraints to be implemented. In the untyped setting, when everything can be anything,
some symbols or data structures tend to percolate into undesirable contexts~\cite{Untagged}. In order to prevent this from happening, some
auxiliary constraints (``\lstinline{absent$^o$}'', ``\lstinline{symbol$^o$}'', etc.) were introduced. These constraints play a role
of a weak dynamic type system, cutting undesirable answers out at runtime. Conversely, in a typed language this work can be
entrusted to the type checker (at the price of enforcing an end user to write properly typed specifications), not only improving the
performance of the system but also reducing the number of constraints which have to be implemented. Finally, it is rather natural
to expect better performance of a statically-typed implementation.

We present an implementation of a set of relational combinators and syntax extensions for
OCaml\footnote{The source code of our implementation is accessible from \url{https://github.com/dboulytchev/OCanren}.},
which, technically speaking, corresponds to $\mu$Kanren~\cite{MicroKanren} with disequality
constraints~\cite{CKanren}. The contribution of our work is as follows:

\begin{enumerate}
\item Our embedding allows an end user to enjoy strong static typing and type inference in relational
specifications; in particular, all type errors are reported at compile time and the types for
all logical variables are inferred from the context.

\item Our implementation is based on the \emph{polymorphic unification}, which, like the polymorphic comparison,
can be used for arbitrary types. The implementation of polymorphic unification uses unsafe features and
relies on the intrinsic knowledge of the runtime representation of values; we show, however, that this does not
compromise type safety.

\item We describe a uniform and scalable pattern for using types for relational programming, which
helps in converting typed data to and from the relational domain. With this pattern, only one
generic feature (``\lstinline{Functor}'') is needed, and thus virtually any generic
framework for OCaml can be used. Although being rather a pragmatic observation, this pattern, we
believe, would lead to a more regular and easy way to maintain relational specifications.

\item We provide a simplified way to integrate relational and functional code. Our approach utilizes
a well-known pattern~\cite{Unparsing, DoWeNeed} for variadic function implementation and makes it
possible to hide the reification of the answers phase from an end user.
\end{enumerate}

The rest of the paper is organized as follows: in the next section we provide a short overview of the related
works. Then we briefly introduce \miniKanren in
its original form to establish some notions; we do not intend to describe the language in its full bloom (interested readers can
refer to~\cite{TRS}). In Section~\ref{sec:goals} we describe some basic constructs behind a \miniKanren implementation, this time
in OCaml. In Section~\ref{sec:unification} we discuss polymorphic unification, and show that unification with
triangular substitution respects typing. Then we present our approach to handle user-defined types by injecting them
into the logic domain, and describe a convenient generic programming pattern, which can be used to implement the conversions from/to logic
domain. We also consider a simple approach and a more elaborate and efficient tagless variant (see Section~\ref{sec:injection}).
Section~\ref{sec:reification} describes top-level primitives and addresses the problem of relational and functional code integration.
Then, in Section~\ref{sec:examples} we present a set of relational examples, written with the aid of our
library. Section~\ref{sec:evaluation} contains the results of a performance evaluation and a comparison of our implementations
with existing implementation for Scheme. The final section concludes.

The authors would like to express a deep gratitude to the anonymous rewievers for their numerous constructive comments, Michael Ballantyne, Greg Rosenblatt, 
and the other attendees of the miniKanren Google Hangouts, who helped the authors in understanding the subtleties of the original miniKanren
implementation, Daniel Friedman for his remarks on the preliminary version of the paper, and William Byrd for all his help and support, which cannot be
overestimated.

\section{Related Works}
\label{sec:relworks}

There is a predictable difficulty in implementing \miniKanren for a strongly typed language.
Designed in the metaprogramming-friendly and dynamically typed realm of Scheme/Racket, the original
\miniKanren implementation pays very little attention to what has a significant importance in (specifically)
ML or Haskell. In particular, one of the capstone constructs of \miniKanren~--- unification~--- has to work for
different data structures, which may have types different beyond parametricity.

There are a few ways to overcome this problem. The first one is simply to follow the untyped paradigm and
provide unification for some concrete type, rich enough to represent any reasonable data structures.
Some Haskell \miniKanren libraries\footnote{\url{https://github.com/JaimieMurdock/HK}, \url{https://github.com/rntz/ukanren}}
as well as the previous OCaml implementation\footnote{\url{https://github.com/lightyang/minikanren-ocaml}} take this approach.
As a result, the original implementation can be retold with all its elegance; the relational specifications, however,
become weakly typed. A similar approach was taken in early works on embedding Prolog into Haskell~\cite{PrologInHaskell}.

Another approach is to utilize \emph{ad hoc} polymorphism and provide a type-specific unification for each ``interesting'' type.
Some \miniKanren implementations, such as Molog\footnote{\url{https://github.com/acfoltzer/Molog}} and
MiniKanrenT\footnote{\url{https://github.com/jvranish/MiniKanrenT}}, both for Haskell, can be mentioned as examples.
While preserving strong typing, this approach requires a lot of ``boilerplate''
code to be written, so some automation --- for example, using 
Template Haskell~\cite{SheardTMH}~---
is desirable. In~\cite{TypedLogicalVariables} a separate type class was introduced to both perform the unification
and detect free logical variables in end-user data structures. The requirement for end user to provide a way to represent
logical variables in custom data structures looks superfluous for us since these logical variables would require proper
handling in the rest of the code outside the logical programming subsystem.

There is, actually, another potential approach, but we do not know if anybody has tried
it: implementing unification for a generic representation of types as sum-of-products and fixpoints of
functors~\cite{InstantGenerics, ALaCarte}. Thus, unification would work for any type for which a representation
is provided. We believe that implementing this representation would require less boilerplate code to be written.

As follows from this exposition, a typed embedding of \miniKanren in OCaml can be done with
a combination of datatype-generic programming~\cite{DGP} and \emph{ad hoc} polymorphism. There are 
a number of generic frameworks for OCaml (for example,~\cite{Deriving}). On the other hand, the support
for \emph{ad hoc} polymorphism in OCaml is weak; there is nothing comparable in power to Haskell
type classes, and even though sometimes the object-oriented layer of the language can be used to mimic
desirable behavior, the result, as a rule, is far from satisfactory. Existing proposals for \emph{ad hoc} polymorphism (for example,
modular implicits~\cite{Implicits}) require patching the compiler, which we want to avoid. Therefore, we 
take a different approach, implementing polymorphic unification once and for all logical types~--- a purely \emph{ad hoc} 
approach, since the features which would provide a less \emph{ad hoc} solution are not yet well integrated into the language. To deal
with user-defined types in the relational subsystem, we propose to use their logical representations (see Section~\ref{sec:injection}), 
which free an end user from the burden of maintaining logical variables, and we use generic programming to build conversions from and to logical
representations in a systematic manner.

\section{\miniKanren~--- a Short Presentation}
\label{sec:demo}

In this section we briefly describe \miniKanren in its original form, using a canonical example.
\miniKanren is organized as a set of combinators and macros for Scheme/Racket, designated to describe
a search for the solution of a certain \emph{goal}. There are four domain-specific constructs
to build \emph{goals}:

\begin{itemize}
\item Syntactic unification~\cite{Unification} in the form \lstinline[language=scheme]{(== $t_1$ $t_2$)}, where $t_1$, $t_2$ are
some \emph{terms}; unification establishes a syntactic basis for all other goals. If there is a unifier for
two given terms, the goal is considered satisfied, a most general unifier is kept as a partial solution, and the execution
of current branch continues. Otherwise, the current branch backtracks.

\item Disequality constraint~\cite{CKanren} in the form \lstinline[language=scheme]{($\not\equiv$ $t_1$ $t_2$)}, where
$t_1$, $t_2$ are some terms; a disequality constraint prevents all branches (starting from the current), in which the
specified terms are equal (w.r.t. the search state), from being considered.

\item Conditional construct in the form

\begin{lstlisting}[language=scheme]
   (conde
      [$g^1_1\;\;g^1_2\;\;\dots\;\;g^1_{k_1}$]
      [$g^2_1\;\;g^2_2\;\;\dots\;\;g^2_{k_2}$]
      $\ldots$
      [$g^n_1\;\;g^n_2\;\;\dots\;\;g^n_{k_n}$]
   )
\end{lstlisting}

where each $g^i_j$ is a goal. A \lstinline{conde} goal considers each collection of subgoals, surrounded by square brackets, as
implicit conjunction (so \lstinline[language=scheme]{[$g^i_1\;\;g^i_2\;\;\dots\;\;g^i_{k_i}$] } is considered as a
conjunction of all $g^i_j$) and tries to satisfy each of them independently~--- in other words, operates on them
as a disjunction.

\item Fresh variable introduction construct in the form

\begin{lstlisting}[language=scheme]
   (fresh ($x_1\;\;x_2\;\;\dots\;\;x_k$)
     $g_1$
     $g_2$
     $\ldots$
     $g_n$
   )
\end{lstlisting}

where each $g_i$ is a goal. This form introduces fresh variables $x_1\;\;x_2\;\;\dots\;\;x_k$ and
tries to satisfy the conjunction of all subgoals $g_1\;\;g_2\;\;\dots\;\;g_n$ (these subgoals may contain
introduced fresh variables).
\end{itemize}

As an example consider a list concatenation relation; by a well-established tradition, the names
of relational objects are superscripted by ``$^o$'', hence \lstinline{append$^o$}:

\begin{lstlisting}[mathescape=true,language=scheme,numbers=left,numberstyle=\small,stepnumber=1,numbersep=-5pt]
   (define (append$^o$ x y xy)
      (conde
         [(== '() x) (== y xy)]
         [(fresh (h t ty)
            (== `(,h . ,t) x)
            (== `(,h . ,ty) xy)
            (append$^o$ t y ty))]))
\end{lstlisting}

We interpret the relation ``\lstinline{append$^o$ x y xy}'' as ``the concatenation of \lstinline{x} and \lstinline{y}
equals \lstinline{xy}''. Indeed, if the list \lstinline{x} is empty, then (regardless the content of \lstinline{y}) in order for the relation to hold
the value for \lstinline{xy} should by equal to that of \lstinline{y}~--- hence line 3. Otherwise, \lstinline{x} can be decomposed into the head
\lstinline{h} and the tail \lstinline{t}~--- so we need some fresh variables. We also need the additional variable \lstinline{ty} to designate the list
that is in the relation \lstinline{append$^o$} with \lstinline{t} and \lstinline{y}. Trivial relational reasonings complete the implementation (lines 5-7).

A goal, built with the aid of the aforementioned constructs, can be run by the following primitive:

\begin{lstlisting}[mathescape=true,language=scheme]
   run $n$ ($q_1\dots q_k$) $G$
\end{lstlisting}

Here $n$ is the number of requested answers (or ``*'' for all answers), $q_i$ are fresh query variables, and $G$ is a goal, which can
contain these variables.

The \lstinline{run} construct initiates the search for answers for a given goal and returns a (finite or infinite) list
of answers~--- the bindings for query variables, which represent individual solutions for that query. For example,

\begin{lstlisting}[mathescape=true,language=scheme]
   run 1 (q) (append$^o$ q '(3 4) '(1 2 3 4) )
\end{lstlisting}

\noindent returns a list \lstinline{((1 2))}, which constitutes the answer for a query variable $q$. The process of constructing
the answers from internal data structures of miniKanren interpreter is called \emph{reification}~\cite{WillThesis}.

\section{Streams, States, and Goals}
\label{sec:goals}

This section describes a top-level framework for our implementation. Even though it contains
nothing more than a reiteration of the original implementation~\cite{MicroKanren, CKanren}
in OCaml, we still need some notions to be properly established.

The search itself is implemented using a backtracking lazy stream monad~\cite{KiselyovBacktracking}:

\begin{lstlisting}
   type $\alpha$ stream

   val mplus : $\alpha$ stream -> $\alpha$ stream -> $\alpha$ stream
   val bind  : $\alpha$ stream -> ($\alpha$ -> $\beta$ stream) -> $\beta$ stream
\end{lstlisting}

Monadic primitives describe the shape of the search, and their implementations may
vary in concrete \miniKanren versions.

An essential component of the implementation is a bundle of the following types:

\begin{lstlisting}
   type env         = $\dots$
   type subst       = $\dots$
   type constraints = $\dots$

   type state = env * subst * constraints
\end{lstlisting}

Type \lstinline{state} describes a point in a lazily constructed search tree: type \lstinline{env} corresponds
to an \emph{environment}, which contains some supplementary information (in particular, an environment is needed to
correctly allocate fresh variables), type \lstinline{subst} describes a substitution, which keeps current bindings
for some logical variables, and, finally, type \lstinline{constraints} represents disequality constraints,
which have to be respected. In the simplest case \lstinline{env} is just a counter for the number of the next free
variable, \lstinline{subst} is a map-like structure and \lstinline{constraints} is a list of substitutions.

The next cornerstone element is the \emph{goal} type, which is considered as a transformer of a state into
a lazy stream of states:

\begin{lstlisting}
   type goal = state -> state stream
\end{lstlisting}

In terms of the search, a goal nondeterministically performs one step of the search: for a given
node in a search tree it produces its immediate descendants. On the user level the type \lstinline{goal}
is abstract, and states are completely hidden.

Next to last, there are a number of predefined combinators:

\begin{lstlisting}
   val (&&&)      : goal -> goal -> goal
   val (|||)      : goal -> goal -> goal
   val call_fresh : ($v$ -> goal) -> goal
   ....
\end{lstlisting}

Conjunction ``\lstinline{&&&}'' combines the results of its argument goals using \lstinline{bind},
disjunction ``\lstinline{|||}'' concatenates the results using \lstinline{mplus}, abstraction
primitive \lstinline{call_fresh} takes an abstracted goal and applies it to a freshly created
variable. Type $v$ in the last case designates the type for a fresh variable, which we leave
abstract for now. These combinators serve as the bricks for the implementation of conventional
\miniKanren constructs and syntax extensions (\lstinline{conde}, \lstinline{fresh}, etc.)

Finally, there are two primitive goal constructors:

\begin{lstlisting}
   val (===) : $t$ -> $t$ -> goal
   val (=/=) : $t$ -> $t$ -> goal
\end{lstlisting}

The first one is a unification, while the other is a disequality constraint. Here, we again left
the type of terms $t$ abstract; it will be instantiated later.

In the implementation of \miniKanren both of these goals are implemented using unification~\cite{CKanren}; this
is true for us as well. However, due to a drastic difference between the host languages, the implementation of
efficient polymorphic unification itself leads to a number of tricks with typing and data representation, which are
absent in the original version.

In this setting, the run primitive is represented by the following function:

\begin{lstlisting}
   val run : goal -> state stream
\end{lstlisting}

This function creates an initial state and applies a goal to it. The states in the return stream describe
various solutions for the goal. As the stream is constructed lazily, taking elements one by one makes
the search progress.

To discover concrete answers, the state has to be queried for its contents. As a rule, a few variables
are reified in a state, i.e. their bindings in the corresponding substitution are retrieved.
Disequality constraints for free variables have to be reified additionally (e.g. represented as a list of
``forbidden'' terms). As forbidden terms can contain free variables, the constraint reification
process is recursive.

In our case, the reification is a subtle part, since, as we will see shortly, it can not be implemented in a
type-safe fragment of the language.

\section{Polymorphic Unification}
\label{sec:unification}

We consider it rather natural to employ polymorphic unification in a language already equipped
with polymorphic comparison~--- a convenient, but somewhat disputable\footnote{See, for example,
\url{https://blogs.janestreet.com/the-perils-of-polymorphic-compare}} feature. Like polymorphic comparison,
polymorphic unification performs a traversal of values, exploiting intrinsic knowledge of their runtime
representation. The undeniable benefits of this solution are, first, that in order to perform unification
for user types no ``boilerplate'' code is needed, and, second, that this approach seems to deliver the
most efficient implementation. On the other hand, all the pitfalls of polymorphic comparison are inherited as
well; in particular, polymorphic unification loops for cyclic data structures and does not work for functional
values. Since we generally do not expect any reasonable outcome in these cases, the only remaining problem is that
the compiler is incapable of providing any assistance in identifying and avoiding these cases. Another drawback is that
the implementation of polymorphic unification relies on the runtime representation of values and has to be fixed
every time the representation changes.  Finally, as it is written in an unsafe manner using the \lstinline{Obj} interface,
it has to be carefully developed and tested.

An important difference between polymorphic comparison and unification is that the former only inspects its operands,
while the results of unification are recorded in a substitution (a mapping from logical variables to terms), which
later is used to reify answers. So, generally speaking, we have to show that no ill-typed
terms are constructed as a result. Overall, this property seems to be maintained vacuously, since the very
nature of (syntactic) unification is to detect whether some things can be considered equal. Nevertheless there are
different type systems and different unification implementations; in addition \emph{equal things} can be
\emph{differently typed}, so we provide here a correctness justification for a well-defined abstract case, and will
reuse this conclusion for various concrete cases.

First, we consider three alphabets:

$$
\begin{array}{rcl}
  \tau,\dots&-&\mbox{types}\\
  x^\tau,\dots&-&\mbox{typed logic variables}\\
  C_k^{\tau_1\times\tau_2\times\dots\times\tau_k\to\tau} (k\ge 0),\dots&-&\mbox{typed constructors}
\end{array}
$$

The set of all well-formed typed terms is defined by mutual induction for all types:

$$
t^\tau=x^\tau\mid C_k^{\tau_1\times\tau_2\times\dots\times\tau_k\to\tau}(t^{\tau_1},\,t^{\tau_2},\,\dots,\,t^{\tau_k})
$$

For simplicity from now on we abbreviate the notation $C_k^{\tau_1\times\tau_2\times\dots\times\tau_k\to\tau}(t^{\tau_1},\,t^{\tau_2},\,\dots,\,t^{\tau_k})$ into
$C_k^\tau(t^{\tau_1},\,t^{\tau_2},\,\dots,\,t^{\tau_k})$, keeping in mind that for any concrete constructor and for all its occurrences
in arbitrary terms all its subterms in corresponding positions agree in types.

In this formulation we do not consider any structure over the set of types besides type equality, and we assume all terms are explicitly
attributed to their types at runtime. We employ this property to implement a unification algorithm in regular OCaml, using some
representation for terms and types:

\begin{lstlisting}[mathescape=true]
    val unify : term -> term -> subst option -> subst option
\end{lstlisting}

\noindent where ``\lstinline{term}'' stands for the type representing typed terms, and ``\lstinline{subst}'' stands for the type of
substitution (a partial mapping from logic variables to terms). Unification can fail (hence ``\lstinline{option}'' in the result type),
is performed in the context of existing substitution (hence ``\lstinline{subst}'' in the third argument) and can be
chained (hence ``\lstinline{option}'' in the third argument).

We use exactly the same unification algorithm with triangular substitution as in the reference implementation~\cite{MicroKanren}. We
omit here some not-so-important details (like ``occurs check''), which are kept in the actual implementaion, and refrain from discussing 
the nature and properties of the algorithm
itself (an excellent description, including a certified correctness proof, can be found in~\cite{Kumar}).

The following snippet presents the implementation:

\begin{lstlisting}[mathescape=true,numbers=left,numberstyle=\small,stepnumber=1,numbersep=-5pt]
    let rec unify $t_1^\tau$ $t_2^\tau$ $subst$ =
      let rec walk $s$ $t^\tau$ =
        match $t^\tau$ with
        | $x^\tau$ when $x^\tau\in dom(s)$ -> $\;\;$walk $s$ $(s\;\;x^\tau)$
        | _ -> $t^\tau$
      in
      match $subst$ with
      | None -> None
      | Some $s$ ->
          match walk $s$ $t_1^\tau$, walk $s$ $t_2^\tau$ with
          | $x_1^\tau$, $x_2^\tau$ when $x_1^\tau$ = $x_2^\tau$ -> $subst$
          | $x_1^\tau$, $q_2^\tau$ -> Some ($s\;[x_1^\tau \gets q_2^\tau]$)
          | $q_1^\tau$, $x_2^\tau$ -> Some ($s\;[x_2^\tau \gets q_1^\tau]$)
          | $C^\tau(t_1^{\tau_1},\dots,t_k^{\tau_k})$, $C^\tau(p_1^{\tau_1},\dots,p_k^{\tau_k})$ ->
              unify $t_k^{\tau_k}$ $p_k^{\tau_k}$(.. (unify $t_1^{\tau_1}$ $p_1^{\tau_1}$ $subst$)$..$)
          | $\_$, $\_$ -> None
\end{lstlisting}

We remind the reader that all superscripts correspond to type attributes, which we consider here as
parts of values being manipulated. For example, line 1 means that we apply \lstinline{unify}
to terms $t_1$ and $t_2$, and expect their types to be equal $\tau$. We assume that
at the top level unification is always applied to some terms of the same type and that any
substitution can only be acquired from the empty one by a sequence of unifications.

We are going to show that under these assumptions all type attributes are superfluous~--- they
do not affect the execution of \lstinline{unify} and can be removed. Note that the only place where we
were incapable of providing an explicit type attribute was in line 4, where the result of
substitution application was returned. However, we can prove by induction that any substitution
respects the following property: if a substitution $s$ is defined for a variable $x^\tau$,
then $s\;\;x^\tau$ is attributed with the type $\tau$ (and, consequently, \lstinline{walk $s$ $t^\tau$} always
returns a term of type $\tau$).

Indeed, this property vacuously holds for the empty substitution. Let $s$ be some substitution, for which the
property holds. In line 11 we return an unchanged substitution; in line 10 we perform two calls~---
\lstinline{walk $s$ $t_1^\tau$} and \lstinline{walk $s$ $t_2^\tau$} and match their results. However,
by our induction hypothesis these results are again attributed to the type $\tau$, which justifies the
pattern matching. In line 11 we return the substitution unchanged, in lines 12 and 13 we extend the
existing substitution but preserve the property of interest. Finally, in line 15 we chain a few
applications of \lstinline{unify}; note that, again, all these calls are performed for terms of equal
types, starting from a substitution possessing the property of interest. A simple induction on the
chain length completes the proof.

So, type attributes are inessential~--- they are never analyzed and never restrict pattern matching; hence,
they can be erased completely.
We can notice now that for the representation of terms we can use OCaml's native runtime representation.
It can not be done, however, using regular OCaml~--- we have to utilize the low-level, unsafe interface \lstinline{Obj}.
Note also, we need some way to identify the occurrences of logical variables inside the terms (in the original \miniKanren
implementation the ambiguity between variable and non-variable datum representation is resolved by a convention~--- a luxury
we cannot afford).  We postpone the discussion on this subject until the next section.


We call our implementation \emph{polymorphic}, since at the top level it is defined as

\begin{lstlisting}
   val unify : $\alpha$ -> $\alpha$ -> subst option -> subst option
\end{lstlisting}

The type of substitution is not polymorphic, which means that the compiler completely loses the track
of types of values stored in a substitution. These types are recovered later during the reification-of-answers phase (see Section~\ref{sec:reification}).
Outside the unification the compiler maintains typing, which means that all terms, subterms, and variables agree in their types
in all contexts.

\section{Term Representation and Injection}
\label{sec:injection}

Polymorphic unification, considered in the previous section, works for the values of any type under the assumption that we
are capable of identifying logical variables. The latter depends on the term representation. In the original
implementation all terms are represented as a conventional S-expressions, while logical variables (in a simplest case)~--- as one-element vectors; it's an end user responsibility to respect this convention and refrain from operating
with vectors as a user data.

In our case we want to preserve both strong typing and type inference. Since we have chosen to use polymorphic
unification, it is undesirable to represent logical variables of different types differently (while technically
possible, it would compromise the lightweight approach we used so far). This means that terms with logical
variables have to be typed differently from user-defined data~--- otherwise it would be possible to use
terms in contexts where logical variables are not handled properly. At the same time we do not want term types
to be completely different from user-defined types --- for example, we would like to reuse user-defined constructors, etc.
This considerations boil down to the idea of \emph{logical representation} for a user-defined type. Informally,
a logical representation for the type $t$ is a type $\rho_t$ with a couple of conversion functions:

$$
\begin{array}{rcl}
   \uparrow  \;: t \to \rho_t & - & \mbox{injection}\\
   \downarrow\;: \rho_t \to t & - & \mbox{projection}
\end{array}
$$

The type $\rho_t$ repeats the structure of $t$, but can contain logic variables. So, the injection is total,
while the projection is partial.

It is important to design representations as instances of some generic scheme (otherwise, \miniKanren combinators
could not be properly typed). In addition it is desirable to provide a generic way to build
injection/projection pair in a uniform manner (and, even better, automatically) to lift the burden of
their implementation off the end user shoulders and improve the reliability of the solution. Finally,
the representation must provide a way to detect logic variable occurrences.


In this part we consider two approaches to implementing logical representations.  The first is rather easy to
develop and implement; unfortunately, the implementation demonstrates a poor performance for a number of
important applications. In order to fix this deficiency, we develop a more elaborate technique which
nevertheless reuses some components from the previous one. In Section~\ref{sec:evaluation}
we present the results of performance evaluation for both approaches.

\subsection{Tagged Logical Values}

The first natural solution is to use tagging for representing logical representations.
We introduce the following polymorphic type $[\alpha]$\footnote{In concrete syntax called ``$\alpha\;$\lstinline{logic}''}, which
corresponds to a logical representation of the type $\alpha$:

\begin{lstlisting}
   type $[\alpha]$ = Var of int | Value of $\alpha$
\end{lstlisting}

Informally speaking, any value of type $[\alpha]$ is either a value of type $\alpha$, or a free
logic variable. Note, the constructors of this type cannot be disclosed to an end user, since the only possible way to create a logic variable
should still be by using the ``\lstinline{fresh}'' construct; thus the logic type is abstract in the interface.
Now, we may redefine the signature of abstraction, unification and disequality primitives in the
following manner

\begin{lstlisting}
   val call_fresh : ($[\alpha]$ -> goal) -> goal

   val (===)      : $[\alpha]$ -> $[\alpha]$ -> goal
   val (=/=)      : $[\alpha]$ -> $[\alpha]$ -> goal
\end{lstlisting}

Both unification and disequality constraint would still use the same polymorphic unification; their external, visible type,
however, is restricted to logical types only.

Apart from variables, other logical values can be obtained by injection; conversely, sometimes a logical value can be projected to
a regular one. We supply two basic functions\footnote{In concrete syntax called ``\lstinline{inj}'' and ``\lstinline{prj}''.}
for these purposes

\begin{lstlisting}[mathescape=true]
   val ($\uparrow_\forall$) : $\alpha$ -> $[\alpha]$
   val ($\downarrow_\forall$) : $[\alpha]$ -> $\alpha$

   let ($\uparrow_\forall$) x = Value x
   let ($\downarrow_\forall$) = function Value x -> x | _ -> failwith $\mbox{``not a value''}$
\end{lstlisting}

\noindent which can be used to perform a \emph{shallow} injection/projection. As expected, the injection is total, while the projection is partial.

The shallow pair works well for primitive types; to implement injection/projection for arbitrary types we exploit the
idea of representing regular types as fixed points of functors~\cite{ALaCarte}. For our purposes it is desirable to make
the functors fully polymorphic~--- thus a type, in which we can place a logical variable into arbitrary position,
can be easily manufactured. In addition this approach makes it possible to refactor the existing code to use relational
programming with only minor changes.

To illustrate this approach, we consider an iconic example~--- the list type. Let us have a conventional definition
for a regular polymorphic list in OCaml:

\begin{lstlisting}
   type $\alpha$ list = Nil | Cons of $\alpha$ * $\alpha$ list
\end{lstlisting}

For this type we can only place a logical variable in the position of a list element, but not of the tail, since the tail
always has the type \lstinline{$\alpha$ list}, fixed in the definition of constructor \lstinline{Cons}. In order to create
a full-fledged logical representation, we first have to abstract the type into a fully-polymorphic functor:

\begin{lstlisting}
   type ($\alpha$, $\beta$) $\mathcal L$ = Nil | Cons of $\alpha$ * $\beta$
\end{lstlisting}

Now, the original type can be expressed as

\begin{lstlisting}
   type $\alpha$ list = ($\alpha$, $\alpha$ list) $\mathcal L$
\end{lstlisting}

\noindent and its logical representation~--- as

\begin{lstlisting}
   type $\alpha$ list$^o$ = $[$($[\alpha]$, $\alpha$ list$^o$) $\mathcal L]$
\end{lstlisting}

Moreover, with the aid of conventional functor-specific mapping function

\begin{lstlisting}
   val fmap$_{\mathcal L}$ : ($\alpha$ -> $\alpha^\prime$) -> ($\beta$ -> $\beta^\prime$) -> ($\alpha$, $\beta$) $\mathcal L$ -> ($\alpha^\prime$, $\beta^\prime$) $\mathcal L$
\end{lstlisting}

\noindent both the injection and the projection functions can be implemented:

\begin{lstlisting}[mathescape=true]
   let rec $\uparrow_{\mbox{\texttt{list}}}$ l = $\uparrow_{\forall}$(fmap$_{\mathcal L}$ ($\uparrow_\forall$) $\uparrow_{\mbox{\texttt{list}}}$ l)
   let rec $\downarrow_{\mbox{\texttt{list}}}$ l = fmap$_{\mathcal L}$ ($\downarrow_\forall$) $\downarrow_{\mbox{\texttt{list}}}$ ($\downarrow_\forall$ l)
\end{lstlisting}

As functor-specific mapping functions can be easily written or, better, derived automatically using a number of existing frameworks for
generic programming for OCaml, one can easily provide injection/projection pair for user-defined data types.

We now can address the problem of variable identification during polymorphic unification. As we do not know the types, we cannot discriminate logical
variables by their tags only and, thus, cannot simply use pattern matching. In our implementation we perform a variable test
as follows:

\begin{itemize}
\item in an environment, we additionally keep some unique boxed value~--- the \emph{anchor}~--- created by \lstinline{run} at the moment of initial
state generation; the anchor is inherited unchanged in all derived environments during the search session;
\item we change the logic type definition into

\begin{lstlisting}
   type $[\alpha]$ = Var of int * anchor | Value of $\alpha$
\end{lstlisting}

\noindent making it possible to save in each variable the anchor, inherited from the environment, in which the variable was created;

\item inside the unification, in order to check if we are dealing with a variable, we test the conjunction of the following properties:

  \begin{enumerate}
    \item the scrutinee is boxed;
    \item the scrutinee's tag and layout correspond to those for variables (i.e. the values, created with the constructor \lstinline{Var} of
type \lstinline{[$\alpha$]});
    \item the scrutinee's anchor and the current environment's anchor have equal addresses.
  \end{enumerate}
\end{itemize}

Taking into account that the state type is abstract at the user level, we guarantee that only those variables which were
created during the current run session would pass the test, since the pointer to the anchor is unique among all pointers to a boxed value
and could not be disclosed anywhere but in the variable-creation primitive.

The only thing to describe now is the implementation of the reification stage. The reification is represented by the following
function:

\begin{lstlisting}
   val reify : state -> $[\alpha]$ -> $[\alpha]$
\end{lstlisting}

This function takes a state and a logic value and recursively substitutes all logic variables in that value w.r.t.
the substitution in the state until no occurrences of bound variables are left. Since in our implementation the type of a substitution is
not polymorphic, \lstinline{reify} is also implemented in an unsafe manner. However, it is easy to see that \lstinline{reify}
does not produce ill-typed terms. Indeed, all original types of variables are preserved in a substitution; unification does not
change unified terms, so all terms bound in a substitution are well-typed. Hence, \lstinline{reify} always substitutes
some subterms in a well-typed term with other terms of the corresponding types, which preserves the well-typedness.

In addition to performing substitutions, \lstinline{reify} also reifies disequality constraints. Constraint reification
attaches to each free variable in a reified term a list of reified terms, describing the disequality constraints for that
free variable. Note, disequality can be established only for equally-typed terms, which justifies the type-safety of reification.
Note also, additional care has to be taken to avoid infinite looping, since reification of answers and constraints are
mutually recursive, and the reification of a variable can be potentially invoked from itself due to a chain of disequality
constraints. In the following example

\begin{lstlisting}
   let foo q =
      fresh (r s)
        (q === $\uparrow_{\forall}$ (Some r)) &&&
        (r =/= s) &&&
        (s =/= r)
\end{lstlisting}

\noindent the answer for the variable $q$ will contain a disequality constraint for the variable $r$; the reification of $r$ will in turn
lead to the reification of its own constraint, this time the variable $s$; finally, the reification of $s$ will again invoke the
reification of $r$, etc.

After the reification, the content of a logical value can be inspected via the following function:

\begin{lstlisting}
   val destruct : $[\alpha]$ -> [`Var of int * $[\alpha]$ list | `Value of $\alpha$]
\end{lstlisting}

Constructor \lstinline{`Var} corresponds to a free variable with unique integer identifier and a list of terms,
representing all disequality constraints for this variable.

\subsection{Tagless Logical Values and Type Bookkeeping}

The solution presented in the previous subsection suffers from the following deficiency: in order to perform unification,
we inject terms into the logic domain, making them as twice as large. As a result, this implementation loses to the original one in
terms of performance in many important applications, which compromises the very idea of using OCaml as a host language.

Here we develop an advanced version, which eliminates this penalty. As a first step, let's try to eliminate the tagging with
a drastic measure:

\begin{lstlisting}
   type $[\alpha]$ = $\alpha$
\end{lstlisting}

What consequences would this have? Of course, we would not be able to create logical variables in a conventional way. However,
we still could have a separate type of variables

\begin{lstlisting}
   type var = Var of int * anchor
\end{lstlisting}

\noindent and use \emph{the same} variable test procedure. As the type $[\alpha]$ is abstract, this modification does not change the interface.
As we reuse the variable test, polymorphic unification can continue to work \emph{almost} correctly. The problem is that
now it can introduce the occurrences of free logic variables in non-logical, tagless, data structures. These free logic variables
do not get in the way of unification itself (since it can handle them properly, thanks to the variable test), but they can not
be disclosed to the outer world as is.

Our idea is to use this generally unsound representation for all internal actions, and perform tagging only during the reification
stage. However, this scenario raises the following question: what would the type of \lstinline{reify} be? It can not be simply

\begin{lstlisting}
   val reify : state -> $[\alpha]$ -> $[\alpha]$
\end{lstlisting}

anymore since $[\alpha]$ now equals $\alpha$. We \emph{want}, however, it be something like

\begin{lstlisting}
   val reify : state -> $[\alpha]$ -> $(\mbox{``tagged'' } [\alpha])$
\end{lstlisting}

If $\alpha$ is not a parametric type, we can simply test if the value is a variable, and if yes, tag it with the constructor \lstinline{Var};
we tag it with \lstinline{Value} otherwise, and we're done. This trick, however, would not work for parametric types. Consider, for example,
the reification of a value of type \lstinline{$[[$int$]$ list$]$}. The (hypothetical) approach being described would return a value of
type \lstinline{$(\mbox{``tagged'' }[[$int$]$ list$])$}, i.e. tagged only on the top level; we need to repeat the procedure
recursively. In other words, we need the following (meta) type for the reification primitive:

\begin{lstlisting}
   val reify : state -> $[\alpha]$ -> $\mbox{(``tagged''} [\beta])$
\end{lstlisting}

\noindent where $\beta$ is the result of tagging $\alpha$.

These considerations can be boiled down to the following concrete implementation.

First, we roll back to the initial definition of $[\alpha]$~--- it will play the role of our ``tagged'' type.
We introduce a new, two-parameter type\footnote{In concrete syntax called ``$(\alpha,\;\beta)\;$\lstinline{injected}''.}

\begin{lstlisting}
   type $\{\alpha,\;\beta\}$ = $\alpha$
\end{lstlisting}

Of course, this type is kept abstract at the end-user level. Informally speaking, the type $\{\alpha,\;\beta\}$ designates the
injection of a tagless type $\alpha$ into a tagged type $\beta$; the value itself is kept in the tagless form, but
the tagged type can be used during the reify stage as a constraint, which would allow us to reify a tagless
representation only to a feasible tagged one. In other words, we record the injection steps using the second
type parameter of the type ``\{,\}'', performing the bookkeeping on the type level rather than on the value level.

We introduce the following primitives for the type $\{\alpha,\;\beta\}$:

\begin{lstlisting}
   val lift : $\alpha$ -> $\{\alpha,\;\alpha\}$
   val inj  : $\{\alpha,\;\beta\}$ -> $\{\alpha,\;[\beta]\}$

   let lift x = x
   let inj  x = x
\end{lstlisting}

The function \lstinline{lift} puts a value into the ``bookkeeping injection'' domain for the first time, while
\lstinline{inj} plays the role of the injection itself. Their composition is analogous to what was
called ``$\uparrow_\forall$'' in the previous implementation:

\begin{lstlisting}
   val $\uparrow_\forall$ : $\alpha$ -> $\{\alpha,\;[\alpha]\}$
   let $\uparrow_\forall$ x = inj (lift x)
\end{lstlisting}

In order to deal with parametric types, we can again utilize generic programming. To handle the types with
one parameter, we introduce the following functor:

\begin{lstlisting}
   module FMap (T : sig type $\alpha$ t val fmap : ($\alpha$ -> $\beta$) -> $\alpha$ t -> $\beta$ t end) :
     sig
       val distrib : $\{\alpha,\;\beta\}$ T.t -> $\{\alpha$ T.t, $\beta$ T.t$\}$
     end =
     struct
       let distrib x = x
     end
\end{lstlisting}

Note, that we do not use the function ``\lstinline{T.fmap}'' in the implementation; however, first, we need an inhabitant of the
corresponding type to make sure we are indeed dealing with a functor, and next, we actually will use it in the
implementation of type-specific reification, see below.

In order to handle two-, three-, etc. parameter types we need higher-kinded polymorphism, which is
not supported in a direct form in OCaml. So, unfortunately, we need to introduce separate
functors for the types with two-, three- etc. parameters; existing works on higher-kinded
polymorphism in OCaml~\cite{HKinded} require the similar scaffolding to be erected as a bootstrap step.

Given the functor(s) of the described shape, we can implement logic representations for
all type's constructors. For example, for standard type \lstinline{$\alpha$ option} with two constructors
\lstinline{None} and \lstinline{Some} the implementation looks like as follows:

\begin{lstlisting}
   module FOption = FMap (struct
     type $\alpha$ t = $\alpha$ option
     let fmap = fmap$_{\mbox{\texttt{option}}}$
   end)

   val some : $\{\alpha, \beta\}$ -> $\{\alpha\mbox{\texttt{ option}},\;\beta\mbox{\texttt{ option}}\}$
   val none : unit  -> $\{\alpha\mbox{\texttt{ option}},\;\beta\mbox{\texttt{ option}}\}$

   let some x  = inj (FOption.distrib (Some x))
   let none () = inj (FOption.distrib None)
\end{lstlisting}

In other words, we can in a very systematic manner define logic representations for all constructors
of types of interest. These representations can be used in the relational code, providing a well-bookkept
typing~--- for each logical type we would be able to reconstruct its original, tagless preimage.

With the new implementation, the types of basic goal constructors have to be adjusted:

\begin{lstlisting}
   val (===) : $\{\alpha,\;[\beta]\}$ -> $\{\alpha,\;[\beta]\}$ -> goal
   val (=/=) : $\{\alpha,\;[\beta]\}$ -> $\{\alpha,\;[\beta]\}$ -> goal
\end{lstlisting}

As always, we require both arguments of unification and disequality constraint to be of the same type; in addition
we require the injected part of the type to be logical.

During the reification stage the bindings for the top-level variables, reconstructed using the final
substitution, have to be properly tagged. This process is implemented in a datatype-generic manner as well:
first, we have reifiers for all primitive types:

\begin{lstlisting}
   val reify$_{\mbox{\texttt{int}}}$ : helper -> $\{$int,$[$int$]\}$ -> $[$int$]$
   val reify$_{\mbox{\texttt{string}}}$ : helper -> $\{$string,$[$string$]\}$ -> $[$string$]$
   ...
\end{lstlisting}

and, then, we add the reifier to the output signature in all \lstinline{FMap}-like functors:

\begin{lstlisting}
   val reify: (helper -> $\{\alpha,\;\beta\}$ -> $\beta$) -> helper -> $\{\alpha$ T.t, $[\beta$ T.t$]$ as $\gamma\}$ -> $\gamma$
\end{lstlisting}

Note, since now \lstinline{reify} is a type-specific and, hence, constructed at the user-level, we refrain from passing
it a state (which is inaccessible on the user level). Instead, we wrap all state-specific functionality in
an abstract \lstinline{helper} data type, which encapsulates all state-dependent functionality needed for \lstinline{reify}
to work properly.

\section{Reification and Top-Level Primitives}
\label{sec:reification}

In Section~\ref{sec:goals} we presented a top-level function \lstinline{run}, which
runs a goal and returns a stream of states. To acquire answers to the query,
represented by that goal, its free variables have to be reified in these states, and
we described the reification primitives in Section~\ref{sec:injection}. However,
the states keep answers in an untyped form, and the types of answers are
recovered solely on the basis of the types of variables being reified. So, the
type safety of the reification critically depends on the requirement to
reify each variable only in those states, which are descendants (w.r.t. the search tree)
of the state, in which that variable was created. In this section we describe a set of
top-level primitives, which enforce this requirement.

We provide a set of top-level combinators, which should be used to surround relational code
and perform reification in a transparent manner only in correct states.
We reimplement the top-level primitive \lstinline{run} to take three
arguments. The exact type of \lstinline{run} is rather complex and non-instructive,
so we prefer to describe the typical form of its application:

\begin{lstlisting}[mathescape=true]
   run $\overline{n}$ (fun $l_1\dots l_n$ -> $\;\;G$) (fun $a_1\dots a_n$ -> $\;\;H$)
\end{lstlisting}

Here $\overline{n}$ stands for a \emph{numeral}, which describes the number of
parameters for two other arguments of \lstinline{run}, \mbox{$l_1\dots l_n$}~---
free logical variables, $G$~--- a goal (which can make use of \mbox{$l_1\dots l_n$}),
\mbox{$a_1\dots a_n$}~--- reified answers for \mbox{$l_1\dots l_n$}, respectively, and,
finally, $H$~--- a \emph{handler} (which can make use of \mbox{$a_1\dots a_n$}).

The types of \mbox{$l_1\dots l_n$} are inferred from $G$ and always have a form

\begin{lstlisting}
   $\{\alpha,\;[\beta]\}$
\end{lstlisting}

\noindent since the types of variables can be constrained only in unification or disequality constraints.

The types of \mbox{$a_1\dots a_n$} are inferred from the types of \mbox{$l_1\dots l_n$} and
have the form

\begin{lstlisting}
   $(\alpha,\;\beta)$ reified stream
\end{lstlisting}

\noindent where the type \lstinline{reified}, in turn, is

\begin{lstlisting}
   type ($\alpha$, $\beta$) reified = $<\;$prj : $\alpha$; reify : (helper -> $\{\alpha,\;\beta\}$ -> $\beta$) -> $\beta>$
\end{lstlisting}

Two methods of this type can be used to perform two different styles of reification: first, a value without
free variables can be returned as is (using the method \lstinline{prj} which checks that in the value of
interest no free variables occur, and raises an exception otherwise). If the value contains some free
variables, it has to be properly injected into the logic domain~--- this is what \lstinline{reify} stands
for. It takes as an argument a type-specific tagging function, constructed using generic
primitives described in the previous section.

In other words a user-defined handler takes streams of reified answers for all variables supplied to the top-level
goal. All streams $a_i$ contain coherent elements, so they all have the same length and $n$-th elements of all
streams correspond to the $n$-th answer, produced by the goal $G$.

There are a few predefined numerals for one, two, etc. arguments (called, traditionally,
\lstinline{q}, \lstinline{qr}, \lstinline{qrs} etc.), and a successor function, which
can be applied to existing numeral to increment the number of expected arguments. The
implementation technique generally follows~\cite{Unparsing, DoWeNeed}.

Thus, the search and reification are tightly coupled; it is simply impossible to perform the reification
for arbitrarily-taken state and variable. This solution both guarantees the type safety and frees an end
user from the necessity to call reification primitives manually.

\section{Examples}
\label{sec:examples}

In this section we present some examples of a relational specification, written with the aid of our library.
Besides \miniKanren combinators themselves, our implementation contains two syntax extensions~--- one
for \lstinline{fresh} construct and another for \emph{inverse-$\eta$-delay}~\cite{MicroKanren}, which is
sometimes necessary to delay recursive calls in order to prevent infinite looping. In addition, we included a
small relational library of data structures like lists, numbers, booleans, etc. This library is written
completely on the user level using techniques described in Section~\ref{sec:injection} with no utilization
of any unsafe features. The examples given below illustrate the usage of all these elements as well.

\subsection{List Concatenation and Reversing}

List concatenation and reversing are usually the first relational programs considered, and we do not wish
to deviate from this tradition. We've already considered the implementation of \lstinline{append$^o$} in
original \miniKanren in Section~\ref{sec:demo}. In our case, the implementation looks familiar:

\begin{lstlisting}
   let rec append$^o$ x y xy =
     (x === nil ()) &&& (y === xy) |||
     (fresh (h t)
       (x === h % t)
       (fresh (ty)
         (h % ty === xy)
         (append$^o$ t y ty)
       )
     )

   let rec revers$^o$ a b =
     conde [
       (a === nil ()) &&& (b === nil ());
       (fresh (h t)
         (a === h % t)
         (fresh (a')
            (append$^o$ a' !< h b)
            (revers$^o$ t a')
         )
       )
     ]
\end{lstlisting}

Here we make use of our implementation of relational lists, which provides convenient shortcuts for
standard functional primitives:

\begin{itemize}
  \item ``\lstinline{nil ()}'' corresponds to ``\lstinline{[]}'';
  \item ``\lstinline{h 
  \item ``\lstinline{a 
\end{itemize}

In our implementation the basic \miniKanren primitive ``\lstinline{conde}'' is implemented as a
disjunction of a list of goals, not as a built-in syntax construct. We also make use of explicit
conjunction and disjunction infix operators instead of nested bracketed structures which, we
believe, would look too foreign here.

\subsection{Relational Sorting and Permutations}

For the next example we take list sorting; specifically, we present a sorting for lists of natural numbers
in Peano form since our library already contains built-in support for them. However, our example can be
easily extended for arbitrary (but linearly ordered) types.

List sorting can be implemented in \miniKanren in a variety of ways~--- virtually any existing algorithm can
be rewritten relationally. We, however, try to be as declarative as possible to demonstrate the
advantages of the relational approach. From this standpoint, we can claim that the sorted version of an empty list is an
empty list, and the sorted version of a non-empty list is its smallest element, concatenated with a sorted
version of the list containing all its remaining elements.

The following snippet literally implements this definition:

\begin{lstlisting}
   let rec sort$^o$ x y =
     conde [
       (x === nil ()) &&& (y === nil ());
       fresh (s xs xs')
         (y === s % xs')
         (sort$^o$ xs xs')
         (smallest$^o$ x s xs)
     ]
\end{lstlisting}

The meaning of the expression ``\lstinline{smallest$^o$ x s xs}'' is ``\lstinline{s} is the smallest element of a (non-empty) list \lstinline{x}, and \lstinline{xs} is the
list of all its remaining elements''. Now, \lstinline{smallest$^o$} can be implemented using a case analysis (note, ``\lstinline{l}'' here is a non-empty list):

\begin{lstlisting}
   let rec smallest$^o$ l s l' =
     conde [
       (l === s % nil ()) &&& (l' === nil ());
       fresh (h t s' t' max)
         (l' === max % t')
         (l === h % t)
         (minmax$^o$ h s' s max)
         (smallest$^o$ t s' t')
     ]
\end{lstlisting}

Finally, we implement a relational minimum-maximum calculation
primitive:

\begin{lstlisting}
   let minmax$^o$ a b min max =
     conde [
       (min === a) &&& (max === b) &&& (le$^o$ a b);
       (max === a) &&& (min === b) &&& (gt$^o$ a b)
     ]
\end{lstlisting}

Here ``\lstinline{le$^o$}'' and ``\lstinline{gt$^o$}'' are built-in comparison goals for natural numbers in Peano form.

Having relational \lstinline{sort$^o$}, we can implement sorting for regular integer lists:

\begin{lstlisting}
   let sort l =
     run q (sort$^o$ (inj_nat_list l))
           (fun qs -> from_nat_list ((Stream.hd qs)#prj) )
\end{lstlisting}

Here \lstinline{Stream.hd} is a function which takes a head from a lazy stream of answers,
\lstinline{inj_nat_list}~--- an injection from regular integer lists into logical lists of logical Peano numbers,
\lstinline{from_nat_list}~--- a projection from lists of Peano numbers to lists of integers.

It is interesting, that since \lstinline{sort$^o$} is
relational, it can be used to calculate a list of all \emph{permutations}
for a given list. Indeed, sorting each permutation results in the same list.
So, the problem of finding all permutations can be relationally reformulated into
the problem of finding all lists which are converted by sorting into the given one:

\begin{lstlisting}
let perm l = map (fun a -> from_nat_list a#prj)
  (run q (fun q -> fresh (r)
                     (sort$^o$ (inj_nat_list l) r)
                     (sort$^o$ q r)
         )
         (Stream.take ~n:(fact (length l))))
\end{lstlisting}

Note, for sorting the original list we used exactly the same primitive. Note also,
we requested exactly \lstinline{fact @@ length l} answers; requesting more
would result in an infinite search for non-existing answers.

\subsection{Type Inference for STLC}

Our final example is a type inference for Simply Typed Lambda Calculus~\cite{Lambda}. The problem and
solution themselves are rather textbook examples again~\cite{TRS, WillThesis}; however, with this example
we show once again the utilization of generic programming techniques we described in Section~\ref{sec:injection}.
As a supplementary generic programming library here we used object-oriented generic transformers\footnote{\url{https://github.com/dboulytchev/GT}};
we presume, however, that any other framework could equally be used.

We first describe the type of lambda terms and their logic representation:

\begin{lstlisting}
   module Term = struct
     module T = struct
       @type ('varname, 'self) t =
       | V   of 'varname
       | App of 'self    * 'self
       | Abs of 'varname * 'self
       with gmap

       let fmap f g x = gmap(t) f g x
     end

     include T
     include FMap2(T)

     let v   s   = inj (distrib (V s))
     let app x y = inj (distrib (App (x, y)))
     let abs x y = inj (distrib (Abs (x, y)))
   end
\end{lstlisting}

Now we have to repeat the work for the type of simple types:

\begin{lstlisting}
     module Type = struct
       module T = struct
         @type ('a, 'b) t =
         | P   of 'a
         | Arr of 'b * 'b
         with gmap

         let fmap f g x = gmap(t) f g x
       end

       include T
       include FMap2(T)

       let p   s   = inj (distrib (P s))
       let arr x y = inj (distrib (Arr (x, y)))
     end
\end{lstlisting}

Note, the ``relational'' part is trivial, boilerplate and short (and could even be generated
using a more advanced framework).

The relational type inferencer itself rather resembles the original implementation. The only
difference (besides the syntax) is that instead of data constructors we use their logic
counterparts:

\begin{lstlisting}
   let rec lookup$^o$ a g t =
     fresh (a' t' tl)
       (g === (inj_pair a' t') % tl)
       (conde [
         (a' === a) &&& (t' === t);
         (a' =/= a) &&& (lookup$^o$ a tl t)
       ])

   let infer$^o$ expr typ =
     let rec infer$^o$ gamma expr typ =
       conde [
         fresh (x)
           (expr === v x)
           (lookupo x gamma typ);
         fresh (m n t)
           (expr === app m n)
           (infer$^o$ gamma m (arr t typ))
           (infer$^o$ gamma n t);
         fresh (x l t t')
           (expr === abs x l)
           (typ  === arr t t')
           (infer$^o$ ((inj_pair x t) % gamma) l t')
       ]
     in
     infer$^o$ (nil()) expr typ
\end{lstlisting}


\section{Performance Evaluation}
\label{sec:evaluation}

One of our initial goals was to evaluate what performance impact would choosing OCaml as a host language makes. In addition we spent some
effort in order to implement \miniKanren in an efficient, tagless manner, and, of course, the outcome of this decision also has to be
measured. For comparison we took faster-miniKanren\footnote{\url{https://github.com/webyrd/faster-miniKanren}}~--- a full-fledged
\miniKanren implementation for Scheme/Racket. It turned out that faster-miniKanren implements a number of optimizations~\cite{WillThesis, Optimizations}
to speed up the search; moreover, the search order in our implementation initially was a little bit different. In order to make the comparison
fair, we additionally implemented all these optimizations and adjusted the search order to exactly coincide with
what faster-miniKanren does.

\begin{figure}[t]
\centering
\includegraphics[scale=0.4]{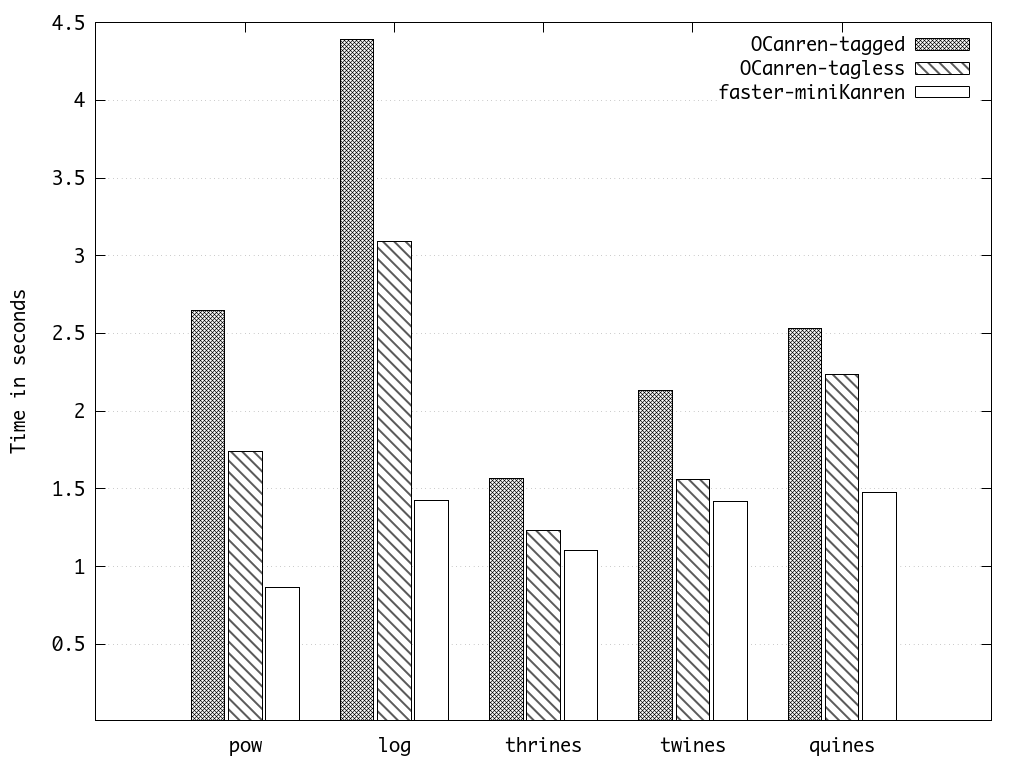}
\caption{The Results of the Performance Evaluation}
\label{eval}
\end{figure}

\FloatBarrier

For the set of benchmarks we took the following problems:

\begin{itemize}
\item \textbf{pow, logo}~--- exponentiation and logarithm for integers in binary form. The concrete tests relationally computed
$3^5$ (which is 243) and $log_3 243$ (which is, conversely, 5). The implementaion was adopted from~\cite{KiselyovArithm}.
\item \textbf{quines, twines, trines}~--- self/co-evaluating program synthesis problems from~\cite{Untagged}. The
concrete tests took the first 100, 15 and 2 answers for these problems respectively.
\end{itemize}


The evaluation was performed on a desktop computer with Intel Core i7-4790K CPU @ 4.00GHz processor and 16GB of memory.
For OCanren \mbox{ocaml-4.04.0+frame_pointer+flambda} was used, for faster-miniKanren~--- Chez~Scheme~9.4.1.
All benchmarks were executed in the natively compiled mode ten times, then average user time was taken. The results of the evaluation
are shown on Figure~\ref{eval}. The whole evaluation repository with all scripts and detailed description is accessible
from \lstinline{GitHub}\footnote{\url{https://github.com/Kakadu/ocanren-perf}}.

The first conclusion, which is rather easy to derive from the results, is that the tagless approach indeed matters. Our initial
implementation did not show essential speedup in comparison even with $\mu$Kanren (and was even \emph{slower} on the logarithm
and permutations benchmarks). The situation was improved drastically, however, when we switched to the tagless version.

Yet, in comparison with faster-miniKanren, our implementation is still lagging behind. We can conclude that the optimizations
used in the Scheme/Racket version, have a different impact in the OCaml case; we save this problem for future research.

\section{Conclusion}

We presented a strongly-typed implementation of \miniKanren for OCaml. Our implementation
passes all tests written for \miniKanren (including those for disequality constraints);
in addition we implemented many interesting relational programs known from
the literature. We claim that our implementation can be used both as a convenient
relational DSL for OCaml and an experimental framework for future research in the area of
relational programming.


\nocite{*}
\bibliographystyle{eptcs}
\bibliography{ocanren}

\end{document}